\documentclass[12pt,nofootinbib]{article}
\usepackage{graphicx}
\usepackage{xcolor}
\usepackage{amsmath,amsfonts}
\usepackage{amsmath}
\usepackage{amsfonts}
\usepackage{amssymb}
\usepackage{hyperref}

\begin{document}

\title{Stability properties of the collective stationary motion of
self-propelling particles with conservative kinematic constraints}
\maketitle
\author{V.I. Ratushnaya$^{1}$, D. Bedeaux$^{1,2}$, V.L. Kulinskii$^{3}$, A.V.
Zvelindovsky$^{4}$\\
\\$^{1}$Colloid and Interface Science group, LIC, Leiden University,
P.O. Box 9502, 2300 RA Leiden, The Netherlands\\ $^{2}$Department of
Chemistry, Norwegian University of Science and Technology, 7491
Trondheim, Norway
\\$^{3}$Department for
Theoretical Physics, Odessa National University, Dvoryanskaya 2,
65026 Odessa, Ukraine\\$^{4}$Department of Physics, Astronomy \&
Mathematics, University of Central Lancashire, Preston PR1 2HE,
United Kingdom}\\


\begin{abstract}
In our previous papers we proposed a continuum model for the
dynamics of the systems of self-propelling particles with
conservative kinematic constraints on the velocities. We have
determined a class of stationary solutions of this hydrodynamic
model and have shown that two types of stationary flow, linear and
radially symmetric (vortical) flow, are possible. In this paper we
consider the stability properties of these stationary flows. We
show, using a linear stability analysis, that the linear solutions
are neutrally stable with respect to the imposed velocity and
density perturbations. A similar analysis of the stability of the
vortical solution is found to be not conclusive.
\end{abstract}

\section{Introduction}

The dynamics of the systems of self-propelling particles (SPP) is of
a great interest for physicists as well as for biologists because of
the complex and fascinating phenomenon of the emergence of the
ordered motion. In nature these systems are represented by flocks of
birds, schools of fishes, groups of bacteria etc.
\cite{Camazine}-\cite{Parrish}. From the physical point of view many
aspects of the observed non-equilibrium phase transition from
disordered to ordered motion are to a large extent still an open
problem.

The first numerical model simulating the behavior of the SPP was
proposed by T. Vicsek at al. \cite{cvaprl1995}. Their model is based
on a kinematic rule imposed on the orientations of the velocities of
the self-propelling particles. At a low noise amplitude and a high
density it was shown that the system undergoes the transition from
the disordered state to coherent motion. The nature of the
transition is not established yet. T. Vicsek's investigation shows
that the occurring transition is of the second order. Several
extensions of T. Vicsek's model have been proposed, which consider
particles with varying velocities, different types of noise, and
including external force fields and/or interparticle attractive and
repulsive forces \cite{Czirok1996}-\cite{Gregoire2004}. The
simulations performed in \cite{Gregoire2004} show the discontinuous
nature of the transition in the T.Vicsek's model.

Properties of T. Vicsek's model were also investigated analytically.
In \cite{Tanner 2005} the spontaneous emergence of ordered motion
has been studied in terms of so-called control laws, using graph
theory. Generalizations of the control laws were considered in \cite%
{Jad2003,Sepulchre2005}. In \cite{Sepulchre2005} it was shown that
the organized motion of SPP with the control laws depending on the
relative orientations of the velocities and relative spacing, can be
of two types only: parallel and circular motion. The stability
properties of these discrete updating rules (including the
T.Vicsek's model) and the dynamics they describe were considered
using Lyapunov theory in \cite{Tanner 2005,Jad2003,Gazi}.

In our first paper \cite{usepll2005} we constructed a hydrodynamic
model for the system of self-propelling particles with conservative
kinematic constraints, which can be considered as a continuum
analogue of the discrete dynamic automaton proposed by T. Vicsek et
al.

Based on the conservation of the kinetic energy and the number of
particles our model is represented by the following equations:
\begin{eqnarray}
\frac{d\mathbf{v}\left( \mathbf{r},t\right) }{dt} &=&\boldsymbol{\omega }%
\left( \mathbf{r},t\right) \times \mathbf{v}\left(
\mathbf{r},t\right) ,
\label{vel} \\
\frac{\partial n}{\partial t}+\nabla \cdot \left( n\left( \mathbf{r}%
,t\right) \mathbf{v}\left( \mathbf{r},t\right) \right)  &=&0,
\label{den}
\end{eqnarray}%
where $\mathbf{v}\left( \mathbf{r},t\right) $ and $n\left( \mathbf{r}%
,t\right) $ are the velocity and the density fields respectively and $%
\boldsymbol{\omega }\left( \mathbf{r},t\right) $ is an angular
velocity field which takes into account the non-potential character
of the interactions between the particles. We modeled this field as
follows:
\begin{eqnarray}
\boldsymbol{\omega }\left( \mathbf{r},t\right)  &=&\int K_{1}\left( \mathbf{r%
}-\mathbf{r^{\prime }}\right) \,n\left( \mathbf{r^{\prime }},t\right) %
\mathop{\rm rot}\nolimits\mathbf{v}\left( \mathbf{r^{\prime }},t\right) \,d%
\mathbf{r^{\prime }}+  \notag  \label{omega} \\
&&\int K_{2}\left( \mathbf{r}-\mathbf{r^{\prime }}\right) \nabla
n\left(
\mathbf{r^{\prime }},t\right) \times \mathbf{v}\left( \mathbf{r^{\prime }}%
,t\right) d\mathbf{r^{\prime }},
\end{eqnarray}%
where $K_{1,2}\left( \mathbf{r}-\mathbf{r^{\prime }}\right) $ are
the averaging kernels. In particular we considered a simple case of
averaging kernels:
\begin{equation}
K_{i}\left( \mathbf{r}-\mathbf{r^{\prime }}\right) =s_{i}\,\delta
\left(
\mathbf{r}-\mathbf{r^{\prime }}\right) ,\text{where}\quad i=1\,\,\text{or}%
\,\,2.
\end{equation}%
We call this the local hydrodynamic model (LHM). In this case Eq.~%
\eqref{omega} reduces to
\begin{equation}
\boldsymbol{\omega }\left( \mathbf{r},t\right) =s_{1}\,n\left( \mathbf{r}%
,t\right) \mathop{\rm rot}\nolimits\mathbf{v}\left(
\mathbf{r},t\right)
+s_{2}\nabla n\left( \mathbf{r},t\right) \times \mathbf{v}\left( \mathbf{r}%
,t\right) ,
\end{equation}%
where
\begin{equation}
s_{i}=\int \,K_{i}\left( \mathbf{r}\right) d\mathbf{r}.
\end{equation}%
In our second article \cite{usphysica} we have shown that the only
regimes of the stationary planar motion in our model are either of
translational or axial symmetry. In this respect our continuum model
gives results similar with those obtained in the discrete model of
T.Vicsek \cite{cvaprl1995,Czirok1996}.

In this paper we investigate the stability of the obtained regimes
of motion with respect to small perturbations. In the next section
we consider the stability of the planar stationary linear flow with
respect to the velocity perturbation directed along the stationary
flow and perpendicular to the flow. We show that in both cases the
evolution of the perturbations has an oscillatory behavior, which
means that they neither grow nor decay with time. This can be
interpreted as neutral stability \cite{arnold} of the corresponding
stationary flow. Also the external pressure term $-\nabla p/n$ can
be included into Eq.~\eqref{vel} in order to account for potential
external forces. In such a case with $s_{2}=0$ there exists the
special case of the incompressible flows, $n = const$, when the
equations of motion coincide with that for potential flow of ideal
fluids. As is known \cite{arnold} such motion in $2D$ geometry is
stable in the Lyapunov sense.

In the third section we consider the stability of the planar
stationary radially symmetric (vortical) motion of SPP with constant
velocity and the density. We find that in this case the linear
analysis does not lead to a conclusive answer about the stability of
the solution.

\section{Stability of planar stationary linear flow in the local
hydrodynamic model}

\subsection{Stability with respect to a velocity perturbation along the flow}

In this section we consider the stability properties of planar
stationary linear flow for the local hydrodynamic model with
$s_{2}=0$, which we further call local hydrodynamic model 1 (LHM1).
At the end of the section we will shortly discuss how these results
extend to the local hydrodynamic models with $s_{1}=0$ and
$s_{1}=s_{2}$. For LHM1 the stationary linear flow is given by
\begin{equation}
\mathbf{v}_{0}\left( \mathbf{r}\right) =v_{0}\,\mathbf{e}_{x}\text{
\ \ and \ \ }n_{0}\left( \mathbf{r}\right) =n_{0},  \label{linstat}
\end{equation}%
where $v_{0}$ and $n_{0}$\ are constants.

\noindent We consider velocity and the density perturbations of the
following form:
\begin{equation}
\mathbf{v}_{1}\left( \mathbf{r},t\right) =v_{0}\,A_{||}\,e^{i\,\mathbf{k}%
\cdot \mathbf{r}}\,e^{\alpha _{||}t}\,\mathbf{e}_{x}\quad \text{\ \
\ and \ \ \ \ }n_{1}\left( \mathbf{r},t\right)
=n_{0}\,B_{||}\,e^{i\,\mathbf{k}\cdot \mathbf{r}}\,e^{\alpha
_{||}t},  \label{pert}
\end{equation}%
The velocity perturbation chosen is directed along the stationary
linear
flow. Here $A_{||},B_{||}$ are constants, $\mathbf{k}=k_{x}\mathbf{e}%
_{x}+k_{y}\mathbf{e}_{y}$ is the wave vector and $\alpha _{||}$ is
an exponent, which determines the time evolution of the
perturbation.

Substituting the solution $\mathbf{v}\left( \mathbf{r},t\right) =\mathbf{v}%
_{0}+\mathbf{v}_{1}\left( \mathbf{r},t\right) ,\,n\left(
\mathbf{r},t\right) =n_{0}+n_{1}\left( \mathbf{r},t\right) $ into
Eqs.~\eqref{vel}-\eqref{den} \noindent we obtain the linearized
system of equations:
\begin{eqnarray}
\frac{\partial \mathbf{v}_{1}}{\partial t}+\left(
\mathbf{v}_{0}\cdot \nabla
\right) \mathbf{v}_{1} &=&s_{1}\,n_{0}\left( \mathop{\rm rot}\nolimits%
\mathbf{v}_{1}\right) \times \mathbf{v}_{0},  \label{linsys1} \\
\frac{\partial n_{1}}{\partial t}+\nabla \cdot \left( n_{0}\mathbf{v}%
_{1}\right) +\nabla \cdot \left( n_{1}\mathbf{v}_{0}\right)  &=&0.
\label{linsys2}
\end{eqnarray}%
For the perturbation \eqref{pert} this system reduces to
\begin{eqnarray}
\frac{\partial v_{1}}{\partial t}+v_{0}\,\frac{\partial
v_{1}}{\partial x}
&=&0,  \label{vstab1} \\
\frac{\partial v_{1}}{\partial y} &=&0,  \label{vstab2} \\
\frac{\partial n_{1}}{\partial t}+v_{0}\,\frac{\partial n_{1}}{\partial x}%
+n_{0}\frac{\partial v_{1}}{\partial x} &=&0.  \label{cestab}
\end{eqnarray}%
Using Eq.~\eqref{pert} one may obtain the relation between $\alpha
_{||}$ and the wave number. From Eq.~\eqref{vstab1} it follows that
\begin{equation}
\alpha _{||}=-ik_{x}v_{0}
\end{equation}%
whereas from the linearized continuity equation \eqref{cestab} we
have
\begin{equation*}
\alpha _{||}=-ik_{x}v_{0}\,\frac{\left( A_{||}+B_{||}\right)
}{B_{||}}.
\end{equation*}%
Both the equalities are satisfied only in the case when $A_{||}=0$.

Thus, in the linear stability analysis with respect to small
deviations of the velocity and density fields, we obtain the
following perturbed solution
\begin{equation}
\mathbf{v}=v_{0}\,\mathbf{e}_{x},\quad n=n_{0}\left[ 1+B_{||}\,e^{ik_{y}y}%
\,e^{ik_{x}\left( x-v_{0}t\right) }\right] .
\end{equation}%
Taking the real part of the density perturbation we have
\begin{equation}
\mathbf{v}=v_{0}\,\mathbf{e}_{x},\quad n=n_{0}\left[ 1+B_{||}\cos
\left( \mathbf{k}\cdot \mathbf{r}-k_{x}v_{0}t\right) \right] \,.
\label{persol}
\end{equation}%
The corresponding density field is shown on Fig1.

\begin{figure}[tbh]
\centering
\includegraphics[angle=-90,scale=0.28]{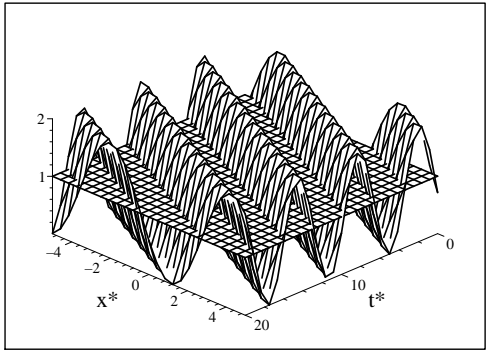}
\caption{The total density field $n\left( \mathbf{r},t\right)
/n_{0}$ and the stationary solution $n/n_{0}=1$ as a function of
$x^{*}=k_{x}x$ and $ t^{*}=k_{x}v_{0}t$ for $k_{y}=0.$}
\end{figure}

\noindent This flow Eq.~\eqref{persol} should satisfy the linearized
system of the constraints (conservation of the kinetic energy and
the number of particles) which are imposed on any solution of our
model. This implies that the following conditions must be fulfilled:
\begin{eqnarray}
\int n_{1}\left( \mathbf{r},t\right) d\mathbf{r} &=&0,  \label{cond1} \\
\int \left[ 2n_{0}\left( \mathbf{v}_{0}\cdot \mathbf{v}_{1}\left( \mathbf{r}%
,t\right) \right) +n_{1}\left( \mathbf{r},t\right) v_{0}^{2}\right] d\mathbf{%
r} &=&0.  \label{cond2}
\end{eqnarray}%
Since $\mathbf{v}_{1}\left( \mathbf{r},t\right) =\mathbf{0}$ both
conditions reduce to
\begin{equation}
\int n_{1}\left( \mathbf{r},t\right) \,d\mathbf{r}=n_{0}\,B_{||}\int
e^{ik_{y}y}dy\int e^{ik_{x}\left( x-v_{0}t\right) }dx=0.  \label{19}
\end{equation}%
If one integrates Eq. \ref{19} over the period of the integrand, one
may can that this condition is fulfilled.

The obtained perturbed flow is an oscillatory field (perturbation
oscillates with a frequency $\alpha_{||}$ as $t\rightarrow\infty$)
which means that the corresponding stationary solution is neither
stable nor unstable within the first order perturbation theory. In
other words we may conclude that in our local hydrodynamic model the
stationary linear flow is neutrally stable with respect to a small
density field perturbations.

The stability analysis of the other possible hydrodynamic models with $%
s_{1}=0$ or $s_{1}=s_{2}$ gives qualitatively similar result.

\subsection{Stability with respect to a velocity perturbation perpendicular
to the flow}

In this section we investigate the stability properties of the
stationary linear flow in the LHM1, Eq.~\eqref{linstat}, with
respect to a velocity perturbation normal to the stationary flow. We
consider only a velocity
perturbation, which we take in the form of a plane wave:%
\begin{equation}
\mathbf{v}_{1}=v_{0}\,A_{\perp }\,e^{i\mathbf{k}\cdot
\mathbf{r}}e^{\alpha _{\perp }t}\mathbf{e}_{y},\quad n_{1}=0,
\end{equation}%
where $A_{\perp }$ is a constant, $\mathbf{k}$\ is a wave vector and
the exponent $\alpha _{\perp }$ describes the time evolution of the
perturbation.

Substituting the perturbation in the linearised equations \eqref{linsys1}-%
\eqref{linsys2}  it follows that
\begin{eqnarray}
\frac{\partial v_{1}}{\partial t}+v_{0}\left( 1-s_{1}n_{0}\right) \frac{%
\partial v_{1}}{\partial x} &=&0,  \label{velper} \\
\frac{\partial v_{1}}{\partial y} &=&0\quad \text{and}\quad k_{y}=0,
\end{eqnarray}%
which imply that
\begin{equation}
\alpha _{\perp }=ik_{x}v_{0}\left( s_{1}n_{0}-1\right) .
\label{alpha}
\end{equation}%
Thus the time evolution of the perturbed velocity field is
determined by the purely imaginary exponent in Eq.~\eqref{alpha}:
\begin{equation}
\mathbf{v}=\mathbf{v}_{0}+\mathbf{v}_{1}\left( x,t\right)
=v_{0}\left[
\mathbf{e}_{x}+A_{\perp }e^{ik_{x}\left( x+\mathbb{V}\,t\right) }\mathbf{e}%
_{y}\right] ,\quad n=n_{0}  \label{solper}
\end{equation}%
where the "phase speed" is given by
\begin{equation*}
\mathbb{V}=v_{0}\left( s_{1}n_{0}-1\right) .
\end{equation*}%
Taking the real part of the velocity perturbation we obtain as final
result:
\begin{equation}
\mathbf{v}=v_{0}\left[ \mathbf{e}_{x}+A_{\perp }\cos \left[ k_{x}\left( x+%
\mathbb{V}t\right) \right] \mathbf{e}_{y}\right] ,\quad n=n_{0}.
\end{equation}%
The corresponding velocity profile is shown in Fig.2.

\begin{figure}[tbh]
\centering
\includegraphics[angle=-90,scale=0.3]{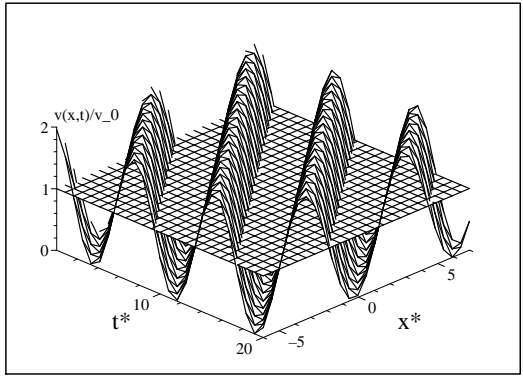}
\caption{The total velocity field $v\left( x,t\right) /v_{0}$ and
the stationary velocity field $v_{0}/v_{0}=1$ as a function of
$x^{*}=k_{x}x$ and $t^{*}=k_{x}\mathbb{V}t$.} \label{velocity}
\end{figure}

Since the velocity perturbation was taken to be normal to the
unperturbed field and $n_{1}=0$, both of the constraints of the
constancy of the kinetic energy and the number of particles,
Eqs.~\eqref{cond1}-\eqref{cond2}, are satisfied.

As one may see the time dependent part of the velocity perturbation
is a finite oscillatory function which means that the corresponding
stationary solution is neutrally stable.

As in the previous section the stability analysis of the other
possible hydrodynamic models with $s_{1}=0$ or $s_{1}=s_{2}$ gives
qualitatively similar result.

\section{Stability of stationary vortical flow with constant velocity and
density in the local hydrodynamic model}

As we have shown in our previous article \cite{usphysica} there are
two classes of the stationary flows in the LHM, linear and radially
symmetric or vortical flow.

The stationary vortical solution of the LHM1 ($s_{2}=0$) is given by $%
\mathbf{v}_{0}\left( \mathbf{r}\right) =v_{\varphi }\left( r\right) \,%
\mathbf{e}_{\varphi },\,n_{0}\left( \mathbf{r}\right) =n_{0}\left( r\right) $%
, \cite{usepll2005}, where
\begin{equation}
v_{\varphi }\left( r\right) =\frac{C_{st}}{2\pi r}\exp \left[
s_{1}\int\limits_{r_{0}}^{r}\frac{dr^{\prime }}{r^{\prime
}\,n_{0}\left( r^{\prime }\right) }\right] .  \label{vvortstat}
\end{equation}%
Here $r_{0}$ is a cut-off radius of the vortex core and the constant
$C_{st}$ is detemined by the circulation of the core
\begin{equation}
\oint_{r=r_{0}}\mathbf{v}d\mathbf{l}=C_{st}.
\end{equation}%
\newline
We consider small perturbations $\mathbf{v}_{1}\left( r,\varphi
,t\right) $\ of the velocity field and $n_{1}\left( r,\varphi
,t\right) $\ of the density field. The linearized system in the LHM1
is then given by
\begin{eqnarray}
\frac{\partial \mathbf{v}_{1}}{\partial t}+\left(
\mathbf{v}_{1}\cdot \nabla
\right) \mathbf{v}_{0}+\left( \mathbf{v}_{0}\cdot \nabla \right) \mathbf{v}%
_{1} &=&s_{1}n_{0}\left[ \left( \mathop{\rm rot}\nolimits\mathbf{v}%
_{1}\right) \times \mathbf{v}_{0}+\left( \mathop{\rm rot}\nolimits\mathbf{v}%
_{0}\right) \times \mathbf{v}_{1}\right]   \notag \\
&&+s_{1}n_{1}\left( \mathop{\rm rot}\nolimits\mathbf{v}_{0}\right)
\times
\mathbf{v}_{0}, \\
\frac{\partial \,n_{1}}{\partial t}+\nabla \cdot \left( n_{0}\mathbf{v}%
_{1}\right) +\nabla \cdot \left( n_{1}\mathbf{v}_{0}\right)  &=&0.
\end{eqnarray}

In this section we consider the stability of a particular class of
stationary vortical flow for which the density is constant and given by $%
n_{0}=1/s_{1}$. Substitution in Eq.~\eqref{vvortstat} results in a
constant velocity field $\mathbf{v}_{0}=v_{\varphi
}\mathbf{e}_{\varphi }=\left(
C_{st}/2\pi r_{0}\right) \mathbf{e}_{\varphi }\equiv C\mathbf{e}_{\varphi }$%
. We write the small perturbation in the general form
\begin{equation}
\mathbf{v}_{1}=a\left( r,\varphi ,t\right) \,\mathbf{e}_{r}+b\left(
r,\varphi ,t\right) \,\mathbf{e}_{\varphi }\text{ \ \ and \ \ }%
\,n_{1}=n_{0}c_{1}\left( r,\varphi ,t\right) .  \label{pertvort}
\end{equation}%
For the projections of the velocity field
$\mathbf{v}=\mathbf{v}_{0}\left( r\right) +\mathbf{v}_{1}\left(
r,\varphi ,t\right) $ together with the continuity equation for the
density field $n=n_{0}+n_{1}\left( r,\varphi ,t\right) $ we have
\begin{eqnarray}
\frac{\partial a}{\partial t}-2\frac{b\,v_{\varphi }}{r}+\frac{v_{\varphi }}{%
r}\,\frac{\partial \,a}{\partial \varphi } &=&-\frac{v_{\varphi
}}{r}\left[
\frac{\partial rb}{\partial r}-\frac{\partial a}{\partial \varphi }\right] -%
\frac{bv_{\varphi }}{r}-c_{1}\frac{v_{\varphi }^{2}}{r},  \label{vvort1} \\
\frac{\partial b}{\partial t}+\frac{v_{\varphi }}{r}\frac{\partial b}{%
\partial \varphi } &=&0,  \label{vvort2} \\
\frac{\partial c_{1}}{\partial t}+\frac{1}{r}\left[ \frac{\partial ra}{%
\partial r}+\frac{\partial b}{\partial \varphi }\right] +\frac{v_{\varphi }}{%
r}\frac{\partial c_{1}}{\partial \varphi } &=&0.  \label{cevort}
\end{eqnarray}%
\newline
In order to simplify the problem we restrict our discussion to the
case with
the radial component of the velocity perturbation being constant, i.e. $%
a\left( r,\varphi ,t\right) =const$.

Then one can transform equations \eqref{vvort1}-\eqref{cevort} into
\begin{eqnarray}
\frac{\partial b}{\partial t}+\frac{v_{\varphi }}{r}\frac{\partial b}{%
\partial \varphi } &=&0,  \label{st1} \\
\frac{\partial b}{\partial r}=-\frac{c_{1}v_{\varphi }}{r}\,, &&
\label{st2}
\\
\frac{\partial c_{1}}{\partial t}+\frac{1}{r}\left( a+\frac{\partial b}{%
\partial \varphi }\right) +\frac{v_{\varphi }}{r}\frac{\partial c_{1}}{%
\partial \varphi } &=&0.  \label{st3}
\end{eqnarray}%
The velocity perturbation must be a periodic functions of the angle
$\varphi $ and can therefore be written as:
\begin{equation}
b\left( r,\varphi ,t\right) =v_{\varphi }B\left( r\right)
e^{im\varphi }e^{\beta t},
\end{equation}%
where $B\left( r\right) $ is a function of $r$, $m$ is an integer
and $\beta $ is a constant factor, which describes the time
evolution of the perturbation, Eq.~\eqref{pertvort}. Substituting
this into Eq.~\eqref{st1} one obtains
\begin{equation}
\beta =-i\,m\frac{v_{\varphi }}{r}  \label{beta}
\end{equation}%
and consequently
\begin{equation}
b\left( r,\varphi ,t\right) =v_{\varphi }B\left( r\right) \exp
\left[ im\left( \varphi -\frac{v_{\varphi }}{r}\,t\right) \right]
\,.  \label{b}
\end{equation}%
From Eq.~\eqref{st2} it follows that
\begin{equation}
c_{1}\left( r,\varphi ,t\right) =-r\left( \frac{\partial B\left( r\right) }{%
\partial r}+im\frac{v_{\varphi }B\left( r\right) }{r^{2}}\,t\right) exp\left[
im\left( \varphi -\frac{v_{\varphi }}{r}\,t\right) \right] .
\label{n1}
\end{equation}%
Substituting this into Eq.~\eqref{st3} we obtain that $a\left(
r,\varphi ,t\right) =0$.

The solutions \eqref{b} and \eqref{n1} satisfy the linearized system
of constraints, Eqs.~\eqref{cond1} and \eqref{cond2}, as one can see
by angular integration.

\noindent Thus, we see that the time evolution of the perturbation Eq.~%
\eqref{pertvort} is determined by the purely imaginary exponent Eq.~%
\eqref{beta}.

\noindent Taking the real part in Eqs.~\eqref{b} and \eqref{n1} we
obtain
\begin{eqnarray}
b\left( r,\varphi ,t\right)  &=&\,v_{\varphi }B\left( r\right) \cos
\left[
m\left( \varphi -\frac{v_{\varphi }}{r}\,t\right) \right] , \\
n_{1}\left( r,\varphi ,t\right)  &=&n_{0}\left\{ \frac{mv_{\varphi
}\,B\left( r\right) }{r}\,t\sin \left[ m\left( \varphi -\frac{v_{\varphi }}{r%
}\,t\right) \right]\right.\nonumber\\
 &&\left.-r\,\frac{\partial B\left( r\right) }{\partial r}\cos %
\left[ m\left( \varphi -\frac{v_{\varphi }}{r}\,t\right) \right]
\right\}
\end{eqnarray}

As a result the whole solution for the velocity and the density
profiles has the following form:
\begin{eqnarray}
\mathbf{v}\left( r,\varphi ,t\right)  &=&v_{\varphi }\left\{
1+B\left(
r\right) \cos \left[ m\left( \varphi -\frac{v_{\varphi }}{r}\,t\right) %
\right] \right\} \,\mathbf{e}_{\varphi }, \\
n\left( r,\varphi ,t\right)  &=&n_{0}\left\{ 1+\frac{mv_{\varphi
}\,B\left(
r\right) }{r}\,t\sin \left[ m\left( \varphi -\frac{v_{\varphi }}{r}%
\,t\right) \right]\right.\nonumber \\
&&\left.-r\,\frac{\partial B\left( r\right) }{\partial r}\cos %
\left[ m\left( \varphi -\frac{v_{\varphi }}{r}\,t\right) \right]
\right\} \,.
\end{eqnarray}%
The velocity field is shown in Fig. 3 for $m=1$ and $r=5\,\rm m$.

\begin{figure}[tbh]
\centering
\includegraphics[angle=-90,scale=0.28]{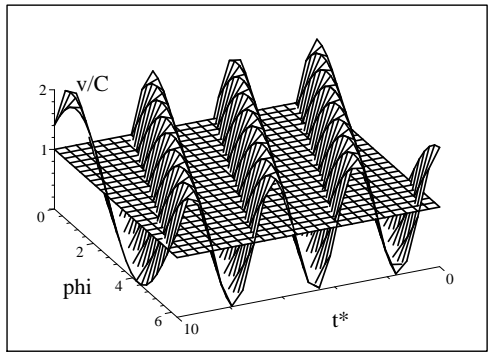}
\caption{The total velocity field $v\left( r,\varphi ,t\right)
/v_{\varphi }$ and $v_{\varphi }\left( r\right) /v_{\varphi }=1$ as
a function of $\varphi $\ and $t^{*}=v_{\varphi}t/r$ for $m=1$ and
$r=5\,\rm m$.}
\end{figure}
Together with the oscillatory contributions we now also have the
contribution proportional to $t$ times an oscillating function. This
does not necessarily mean that the stationary vortical flow is
unstable. The obtained result indicates that the linear analysis
does not give a definite answer regarding the stability of the
stationary flow. The linear analysis does not give the definitive
answer regarding the stability of the stationary flow and further
investigation of higher order terms is required. This is beyond the
scope of the present paper.

\section{Conclusions}

In this paper we considered the stability properties of the planar
stationary flows of the local hydrodynamic model constructed in our
first paper for a system of self-propelling particles
\cite{usepll2005}. These flows are the linear flow and the radially
symmetric flow. Our analysis shows for linear flow, using linear
perturbation theory, that the time evolution of the imposed velocity
and density perturbations are oscillatory. It follows that the
linear flows are neutrally stable. For radially symmetric (vortical)
flow linear perturbation theory does not lead to a conclusive
result. A definitive answer about the nature of the stability can
only be given by considering also higher order terms in the
perturbation expansion. Such an analysis is beyond the scope of the
present paper. Note that such a situation is typical for Hamiltonian
systems which are conservative by definition and therefore do not
display an asymptotic type of stability \cite{arnold}.


\begin{thebibliography}{99}

\bibitem{Camazine}
Camazine S, Deneubourg J-L, Franks N R, Sneyd J, Theraulaz G,
Bonabeau E  2001 \textit{Self-Ogranization in Biological Systems}
(Princeton and Oxford: Princeton University Press)

\bibitem{Parrishbook}
Parrish J K, Hamner W M 1997 \textit{Three dimensional animals
groups} (Cambridge: Cambridge University Press)

\bibitem{Parrish}
Parrish J K, Edelstein-Keshet L 1999 \textit{Science} \textbf{284}
99\\ Parrish J K, Viscido S V, Gr$\mathrm{\ddot{u}}$nbaum D 2002
\textit{Biol. Bull.} \textbf{202} 296

\bibitem{cvaprl1995}
Vicsek T, Czir\'{o}k A, Ben-Jacob E, Cohen I, Shochet O 1995
\textit{Phys. Rev. Lett.} \textbf{75} 1226\\ Czir\'{o}k A, Stanley H
E, Vicsek T 1997 \textit{J. Phys. A: Math. Gen.} \textbf{30} 1375

\bibitem{Czirok1996}
Czir\'{o}k A, Ben-Jacob E, Cohen I and Vicsek T 1996 \textit{Phys.
Rev. E} \textbf{54} 1791

\bibitem{Hubbard2004}
Hubbard S, Babak P, Sigurdsson S Th, Magn\'{u}%
sson K G 2004 \textit{Ecological Modelling} \textbf{174} 359

\bibitem{Gregoire2004}
Gr\'{e}goire G, Chat\'{e} H 2004 \textit{Phys. Rev. Lett.}
\textbf{92} 025702\\ Gr\'{e}goire G, Chat\'{e} H, Tu Y 2003
\textit{Physica D} \textbf{181} 157

\bibitem{Tanner 2005}
Tanner H G, Jadbabaie A, Pappas G J 2005 \textit{Cooperative control
lecture notes in control and information sciences} \textbf{309} 229

\bibitem{Jad2003}
Jadbabaie A, Lin J, Morse A S 2003 \textit{IEEE transactions on
automatic control} \textbf{48} 988

\bibitem{Sepulchre2005}
Sepulchre R, Paley D, Leonard N 2005 \textit{Cooperative control
lecture notes in control and information sciences} \textbf{309} 189

\bibitem{Gazi}
Gazi V, Passino K M 2003 \textit{IEEE transactions on automatic
control} \textbf{48} 692\\ Gazi V, Passino K M 2004 \textit{IEEE
transactions on systems, man, and cybernetics - part B: cybernetics}
\textbf{34} 539

\bibitem{usepll2005}
Kulinskii V, Ratushnaya V, Zvelindovsky A, Bedeaux D 2005
\textit{Europhys. Lett.} \textbf{71} 207

\bibitem{usphysica}
Ratushnaya V, Kulinskii V, Zvelindovsky A, Bedeaux D 2006
\textit{Physica A} \textbf{366} 107

\bibitem{arnold}
Arnold V I 1978 \textit{Mathematical Methods of Classical Mechanics}
(New York, NY: Springer)
\end{thebibliography}
\end{document}